\documentclass[conference]{IEEEtran}
\IEEEoverridecommandlockouts
\usepackage{cite}
\usepackage{amsmath,amssymb,amsfonts}
\usepackage{algorithmic}
\usepackage{graphicx}
\usepackage{textcomp}
\usepackage{xcolor}
\usepackage{braket}
\usepackage{mathtools}
\usepackage{blindtext}
\usepackage{cleveref}
\usepackage{subfig}
\Crefname{figure}{Fig.}{Fig.}

\usepackage{soul}
\sethlcolor{black}
\makeatletter
\newif\if@blind
\if@blind \sethlcolor{black}\else
   \let\hl\relax
\fi

\def\BibTeX{{\rm B\kern-.05em{\sc i\kern-.025em b}\kern-.08em
    T\kern-.1667em\lower.7ex\hbox{E}\kern-.125emX}}
\begin{document}

\title{Clever Design, Unexpected Obstacles: Insights on Implementing a Quantum Boltzmann Machine\\
\thanks{\hl{This work was partially funded by the project PlanQK (01MK20005N) supported by the German Federal Ministry for Economic Affairs and Climate Action.}}
}

\author{\IEEEauthorblockN{\hl{Felix Paul}}
\IEEEauthorblockA{\textit{\hl{StoneOne AG}}\\
\hl{Berlin, Germany} \\
\hl{felix.paul@stoneone.de}}
\and
\IEEEauthorblockN{\hl{Michael Falkenthal}}
\IEEEauthorblockA{\textit{\hl{StoneOne AG}} \\
\hl{Berlin, Germany} \\
\hl{michael.falkenthal@stoneone.de}}
\and
\IEEEauthorblockN{\hl{Sebastian Feld}}
\IEEEauthorblockA{\textit{\hl{Delft University of Technology}}\\
\hl{Delft, The Netherlands}\\
\hl{s.feld@tudelft.nl}}
}
\maketitle

\begin{abstract}
We have implemented a gated-based quantum version of a restricted Boltzmann machine for approximating the ground state of a Pauli-decomposed qubit Hamiltonian.
During the implementation and evaluation, we have noticed a variety of unexpected topics. It starts from limitations due to the structure of the algorithm itself and continues with constraints induced by specific quantum software development kits, which did not (yet) support necessary features for an efficient implementation.
In this paper we systematically summarize our findings and categorize them according to their relevance for the implementation of similar  quantum algorithms. We also discuss the feasibility of executing such implementations on current NISQ devices.
\end{abstract}


\section{Introduction}

The advent of quantum computers accessible to the public has given an immense boost to the development of new quantum algorithms in the last decade.
Many algorithms promise advantages over previously known classical ones, e.g., with regard to a speedup~\cite{Harrow2009} or enhanced solution quality~\cite{Havlicek2019}.
However, to raise these advantages the conceptual algorithms first have to be applied to actual use cases.
Second, they have to be implemented to run on specific quantum computers which are still NISQ devices providing limitations with respect to decoherence time, gate and measurement failure rates~\cite{Preskill2018}.
In practice, this leads to major difficulties when transferring conceptual algorithms into actual implementations.
In order for them to be executable on NISQ-devices, the limitations of the specific quantum hardware must be taken into account accordingly.
For example, the available number of qubits and the fidelity of their states must be considered such that an implemented algorithm can (i) be transferred to the available qubits at all and (ii) be executed by the quantum computer in an amount of time that still guarantees tolerable error rates and, thus, allows to read out meaningful results~\cite{Leymann2020_BitterTruth}.

However, these limitations have a strong influence on whether the theoretically described advantages of an algorithm can be raised at all~\cite{Aaronson2015}.
Therefore, it is valuable to examine algorithms for their practicability on NISQ computers and to recognize limitations early on.
On this basis, it is possible to investigate and elaborate on appropriate mitigation.

In this work, we present details and findings regarding the implementation of a quantum version of a restricted Boltzmann machine (QBM) based on the algorithm introduced by Xia and Kais~\cite{Xia2018}.
We provide insights into what kind of obstacles we encountered when translating a theoretically proposed quantum algorithm into an implementation that can be executed on one of the quantum backends currently available.
Some of these issues are caused by implementing the algorithm within a given software stack or quantum software development kit (quantum SDK).
Others only arise when trying to execute the code on quantum backends, since their properties play a crucial role for gaining meaningful results.

We also discuss aspects regarding the possible solution quality of the chosen model:
There are limitations induced by transferring quantum data, stored within the hardware, into classical data and also some limitations that are induced by the variational model itself and the conditions it is built on.
For some of the mentioned issues we suggest possible approaches, as to how the impact of these may be reduced.
Those recommendations provide guidance for implementing other algorithms which make use of similar concepts.
With this in mind, we classify the encountered problems in terms of how the same or similar ones may arise during the implementation of other algorithms. 
Since numerous algorithms proposed make use of similar subroutines or concepts (e.g., using parametrized rotational gates for variational models), we believe that by systematically classifying common issues, some of these can be managed better within future implementations. 
Thus, we aim on giving an initial list of indicators which need to be considered in order to access the practical feasibility of algorithms proposed.
Since our findings will be explained in the context of a QBM algorithm, we also summarize the key aspects of it which we supplement with exemplary calculations to illustrate relevant procedures in more detail.

The remainder of this paper is structured as follows:
We introduce the essential concepts of the quantum Boltzmann machine algorithm by Xia and Kais~\cite{Xia2018} in Sec. \ref{sec_algo_description}.
We identify and discuss findings and restrictions of the QBM algorithm with respect to currently available quantum computers and software development stacks and development kits in Sec. \ref{sec_insights}.
Finally, we conclude the paper with an outlook on future work in Sec. \ref{sec_conclusion}.

\newpage

\section{Algorithm Details}
\label{sec_algo_description}

The goal of the paper is to highlight practical problems that arise during the implementation of quantum algorithms, which is done on the example of a QBM. For this purpose, we will present the algorithmic details in the following, which will be referred to in~\Cref{sec_insights}.

\begin{figure}[tbp]
\centerline{\includegraphics[scale=.4]{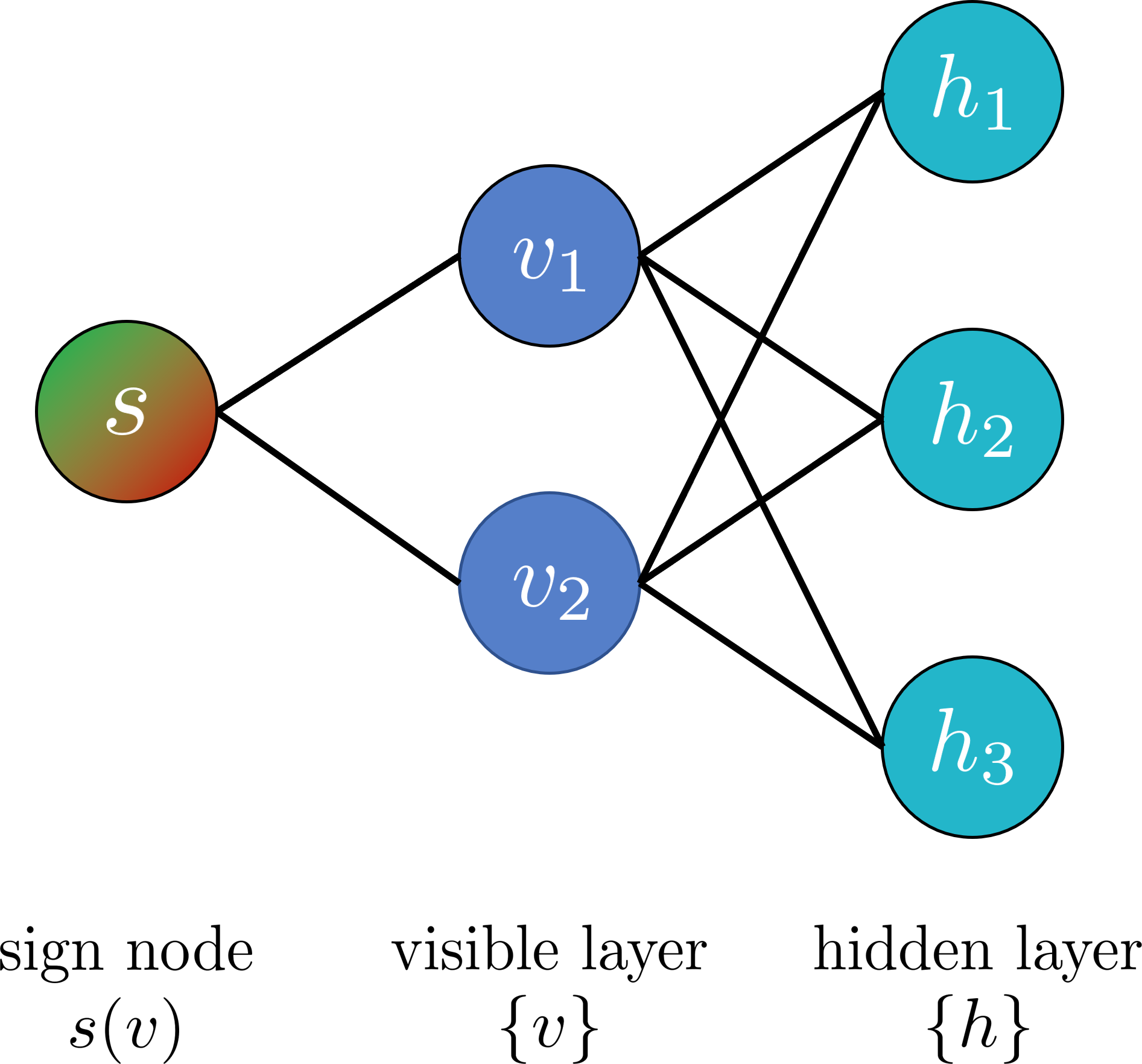}}
\caption{Exemplary network architecture of a $(2,3)$-QBM. Vertices $v_1$ and $v_2$ represent qubits from the visible layer, while $h_1$, $h_2$, and $h_3$ represent qubits from the hidden layer. $s$ corresponds to a sign node realizing relative signs between states of the visible layer. As part of the algorithm, qubits from the visible and hidden layer are entangled with each other via 3-qubit gates.}
\label{fig_qbm_architecture}
\end{figure}

The goal of the approach described by Xia and Kais~\cite{Xia2018} is to approximate the ground state of a given Hamiltonian using a hybrid quantum-classical algorithm.
The utilized QBM is made up of three layers: a visible layer with $n$ qubits, a hidden layer with $m$ qubits, and a classical sign layer realizing relative signs between different states of the visible layer (see \Cref{fig_qbm_architecture}). For explicit values of $n$ and $m$, we will refer to this model as a $(n,m)$-QBM.
The wave function to be generated by the QBM and which should approximate the ground state is
\begin{equation}
    \ket{\psi}=\sum\nolimits_{\lbrace v\rbrace} s(v)\phi(v)\ket{v}~.
\label{eq_trial_wf}
\end{equation}
Here, $v$ denotes a single binary state consisting of $n$ bits and $\lbrace v\rbrace$ represents the set of all possible binary states of the visible layer (e.g., $n=2$ leads to $\lbrace v\rbrace=\lbrace 00, 01, 10, 11\rbrace$).
Furthermore, $\phi(v)$ is the real-valued amplitude associated to state $\ket{v}$, while $s(v)$ is the corresponding value of the sign node.
The probability distribution $p(v)$ of the qubit states of the visible layer is generated by using the QBM, and it determines the amplitudes of the target wave function:
\begin{equation}
    \begin{gathered}
    p(v) = \phi^2(v)= \frac{1}{Z}\sum\nolimits_{\lbrace h\rbrace}\exp\left(E(v,h)\right)~,\\[.1cm] Z=\sum\nolimits_{\lbrace v,h\rbrace}\exp\left(E(v,h)\right)~,
    \end{gathered}
    \label{eq_prob_distribution}
\end{equation}
where $\sum\nolimits_{\lbrace h\rbrace}$ indicates the summation over all configurations of the hidden layer for a fixed configuration of the visible layer.
The "energy" $E(v,h)$ of a given state of the visible and hidden layer depends on real biases $\lbrace a_i\rbrace, \lbrace b_j\rbrace$ and weights $\lbrace w_{ij}\rbrace$, which are all subject to the optimization procedure.
The energy is given by
\begin{align}
    E(v,h)=\sum\nolimits_i a_i v_i + \sum\nolimits_j b_j h_j + \sum\nolimits_{ij}w_{ij}v_i h_j~.
\label{eq_energy}
\end{align}
In~\eqref{eq_energy}, $v_i, h_j\in\lbrace\pm 1\rbrace$ are the $\sigma_z$-eigenvalues of the computational basis states (${\sigma_z\ket{0}=\ket{0}~,~\sigma_z\ket{1}=-\ket{1}}$) for the $i$-th and $j$-th qubit in the visible and hidden layer, respectively.

The sign node is a smooth function that depends on the state of the qubits from the visible layer as well as on parameters $\lbrace c_i\rbrace$ and $d$, which are to be optimized, and it is given by
\begin{align}
    s(v)=\tanh\left(\sum\nolimits_i c_i v_i + d\right)
\label{eq_sign_node}
\end{align}
In fact, the algorithm proposed in~\cite{Xia2018} does not actually generate the probability distribution described in~\eqref{eq_prob_distribution}, but a modified one with an additional regulator $k=\mathcal{O}(\sum\nolimits_{ij}\vert w_{ij}\vert )$. This regulator normalizes all parameters $p\in\lbrace a_i, b_j, w_{ij}\rbrace$ according to $p\rightarrow p/k$ in order to increase the probability for successfully generating the distribution initially wanted (see \Cref{sec_quadratic_terms}).
For simplicity, including the regulator will be omitted in the upcoming sections, but when implementing the algorithm it needs to be considered.

In the following sections we will describe the necessary steps for generating the probability distribution given in~\eqref{eq_prob_distribution} in two steps -- namely generating the \emph{linear} and \emph{quadratic} terms of~\eqref{eq_energy}. After that, the basic optimization procedure and the analytical gradients used in it are introduced.

\subsection{Linear terms}
\label{sec_linear_terms}
The linear terms in~\eqref{eq_energy} are generated by performing $R_y$-rotations on all qubit states from the visible and hidden layer with angle
\begin{align}
    \theta_\ell = 2\arcsin\left(\sqrt{\frac{e^{-p_\ell}}{e^{p_\ell} + e^{-p_\ell}}}\right)~,
\label{eq_single_qubit_angles}
\end{align}
where $p_\ell\in\lbrace a_i, b_j\rbrace$ depends on whether the gate acts on a qubit state from the visible or hidden layer and $\ell$ to be taken from the corresponding index set. To give an example, \eqref{eq_single_qubit_rotation} shows the action on a single $\ket{0}$-state which generates the correct sign within the amplitude of each state:
\begin{align}
    R_y(\theta_\ell)\ket{0}=\frac{1}{\sqrt{e^{p_\ell}+e^{-p_\ell}}}\left(e^{p_\ell/2}\ket{0}+e^{-p_\ell/2}\ket{1}\right)
\label{eq_single_qubit_rotation}
\end{align}
Note, that by rotating the qubit states in the described manner, the probability distribution of~\eqref{eq_prob_distribution} is generated for ${w_{ij}=0, \forall~i,j}$.

\subsection{Quadratic terms}
\label{sec_quadratic_terms}
\begin{figure}[htbp]
\centerline{\includegraphics[scale=1]{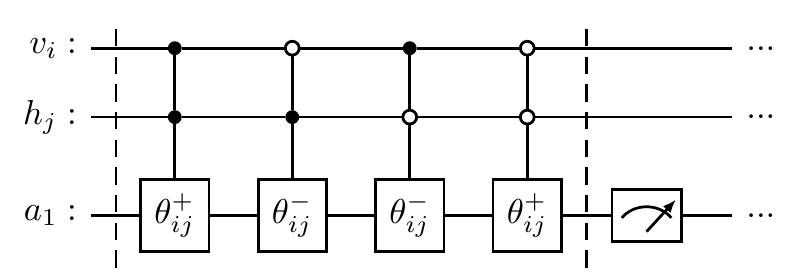}}
\caption{Circuit diagram representation of an entangling layer generating the interaction term between the $i$-th qubit from the visible and the $j$-th qubit from the hidden layer. The gates are $R_y$-gates with the arguments denoted in the box and defined in~\eqref{eq_doubly_rotation_angle}.}
\label{fig_entangling_schema}
\end{figure}

In order to include the interaction term of the $i$-th visible and $j$-th hidden qubit in~\eqref{eq_energy}, a series of four doubly-controlled $R_y$-rotation gates has to act on an ancillary qubit.
The schema of one such entangling layer is depicted in~\Cref{fig_entangling_schema} and shows four 3-qubit gates which all get controlled by one of the four 2-qubit basis states $\lbrace\ket{00}, \ket{01}, \ket{10}, \ket{11}\rbrace$.

For each entangling layer two angles are necessary, which both depend on the weight $w_{ij}$ between qubit $i$ and $j$ in the following way:
\begin{align}
    \theta_{ij}^\pm=2\arcsin\left(\sqrt{\frac{e^{\pm w_{ij}}}{e^{\vert w_{ij}\vert}}}\right)~,
\label{eq_doubly_rotation_angle}
\end{align}
where $\theta_{ij}^+$ is used when the $R_y$-gate is controlled by states with even parity ($\ket{00}$ and $\ket{11}$), while $\theta_{ij}^-$ is accordingly used for control states with odd parity ($\ket{01}$ and $\ket{10}$).
This can better be understood when looking at the energy in~\eqref{eq_energy}: If visible qubit $i$ and hidden qubit $j$ are in the same state (either both $\ket{0}$ or both $\ket{1}$), the product of their $\sigma_z$-eigenvalues is $+1$ (since $(+1)^2=(-1)^2=+1$), whereas them being in different states results in an eigenvalue product of $-1$ (since $(-1)(+1)=(+1)(-1)=-1$).
Now, in order to understand the action of the doubly-controlled $R_y$-gates in more detail, we will look at the action of a single-qubit $R_y$-gate on an ancillary qubit with the angle defined in~\eqref{eq_doubly_rotation_angle} (omitting the $i,j$ indices for simplicity):
\begin{align}
    R_y(\theta^\pm)\ket{0}=\frac{1}{e^{\vert w/2\vert}}\left(\sqrt{e^{\vert w\vert}-e^{\pm w}}\ket{0} + e^{\pm w/2}\ket{1}\right)~.
\label{eq_doubly_controlled_rotation}
\end{align}
Eq.~\eqref{eq_doubly_controlled_rotation} shows that with probability $\sim e^{\pm w}$ the ancillary qubit will be in state $\ket{1}$, thus giving the correct contribution to the energy defined in~\eqref{eq_energy} when measured to be in that particular state.

In order to successfully generate the desired distribution, one ancillary qubit for every combination of qubits from the visible and hidden layer has to be prepared according to the doubly-controlled rotation layer depicted in~\Cref{fig_entangling_schema}, and it has to be measured to be in state $\ket{1}$ directly after the action of the layer.
This results in an additional amount of $nm$ qubits required besides the $n+m$ qubits making up the visible and hidden layer, respectively.
However, the number of required ancillaries could be reduced to $1$ if it is possible to reliably reinitialize it back to the computational ground state $\ket{0}$ after each measurement, resulting in $n+m+1$ required qubits in total.

Furthermore, when looking at the modulus in the normalization factor in~\eqref{eq_doubly_controlled_rotation} it might seem out of place compared to the partition function-like normalization from~\eqref{eq_prob_distribution}.
But after measuring the ancillary qubit to be in state $\ket{1}$, the normalization of the wave function adapts accordingly.

After following the steps described above, the probability distribution of the visible qubit states follows the one given in~\eqref{eq_prob_distribution}, which allows for the sampling of the target wave function defined in~\eqref{eq_trial_wf}.
In order to optimize the involved parameters, the sampled wave function then has to be used for calculating expectation values.
These are necessary for calculating analytic gradients w.r.t. the parameters which will be described in the following.

\subsection{Optimization \& Analytic gradients}
\label{sec_optimization}

It is common to present a problem Hamiltonian in its Pauli-decomposed form according to
\begin{equation}
    H = \sum_{k=1}^N c_k\bigotimes_{i=1}^n P_i^{(k)}~,~P_i^{(k)}\in\lbrace I,\sigma_x, \sigma_y,\sigma_z\rbrace~.
\label{eq_pauli_hamiltonian}
\end{equation}
The Hamiltonian in~\eqref{eq_pauli_hamiltonian} consists of $N$ terms contributing to the sum, each being a so-called Pauli-word or Pauli-string, a tensor product of operators taken from the set of Pauli matrices including the identity, multiplied with a real coefficient $c_k$.
As an example, a Pauli-decomposed Hamiltonian for $n=3$ and $N=2$ might read as
\begin{equation}
    H= 2~\sigma_x\otimes I\otimes \sigma_z - 3~I\otimes \sigma_y\otimes \sigma_y~.
\label{eq_pauli_hamiltonian_example}
\end{equation}
The optimization procedure is straight-forward: For a given set of parameters ${p\in\lbrace a_i, b_j, w_{ij}, c_i, d\rbrace}$, the expectation value $\braket{H}=\braket{\psi\vert H\vert\psi}$ of the Hamiltonian w.r.t. the sampled wave function is the objective function used for the gradient-based optimization of the parameters.
For plain gradient descent with a constant learning rate $\eta$, the parameters in the $k$-th iteration step are adjusted according to
\begin{equation}
    p_{k+1}=p_k-\eta\partial_{p}\braket{H}~.
\label{eq_gradient_descend}
\end{equation}
Given the explicitly known dependence of the amplitude $a(v)\coloneqq s(v)\phi(v)$ on the parameters $p$ and exploiting the fact that the amplitudes are real, the following covariance-like structure can be derived for the derivative of the expectation value (we refer to the supplementary notes of~\cite{Xia2018} for a detailed derivation):
\begin{equation}
    \partial_p\braket{H}=2\braket{E_{loc}D_p}-2\braket{E_{loc}}\braket{D_p}~,
\label{eq_exp_val_derivative}
\end{equation}
with the local energy $E_{loc}(v)$ and the logarithmic amplitude derivative $D_p(v)$ defined as follows
\begin{gather}
    E_{loc}(v)=\frac{\braket{v\vert H\vert\psi}}{a(v)}~,\label{eq_local_energy}\\
    D_p(v)=\partial_p\log\left(a(v)\right)=\frac{\partial_p a(v)}{a(v)}~.
\label{eq_amplitude_derivative}
\end{gather}
For each parameter, the explicit expression of $D_p$ can be worked out analytically to give
\begin{align}
    D_{a_i}(v)&=\frac{1}{2}v_i-\frac{1}{2}\braket{v_i}_{QBM}~,\label{eq_D_a}\\
    D_{b_j}(v)&=\frac{1}{2}\tanh\left(g_j\right)-\frac{1}{2}\braket{h_j}_{QBM}~,\label{eq_D_b}\\
    D_{w_{ij}}(v)&=\frac{1}{2}\tanh\left(g_j\right)v_i-\frac{1}{2}\braket{v_i h_j}_{QBM}~,\label{eq_D_w}\\
    D_{d_i}(v)&=v_i\left(\frac{1}{s(v)}-s(v)\right)~,\label{eq_D_d}\\
    D_c(v)&= \frac{1}{s(v)}-s(v)~,\label{eq_D_c}
\end{align}
with $g_j=\partial E/\partial b_j=h_j+\sum_i w_{ij}v_i$, $v_i$ and $h_j$ again being the corresponding $\sigma_z$-eigenvalues for the $i$-th and $j$-th qubit state in the visible and hidden layer, respectively, and $s(v)$ being the sign node from~\eqref{eq_sign_node}.
Due to the covariant structure of~\eqref{eq_exp_val_derivative}, the constant shifts $\braket{...}_{QBM}$ in~\eqref{eq_D_a}--\eqref{eq_D_w} cancel out and do not have to be calculated.

With the QBM approach just explained, the underlying exponentially growing, complex distribution of $2^n$ states can be generated using quadratically growing resources, namely the circuit width (assuming $m\sim\mathcal{O}(n)$), circuit depth (according to $nm$ entangling layers necessary to generate the quadratic terms), and required parameters.
However, during the investigation and implementation of the approach, several (unexpected) observations have been made that, when applied, diminish the practicability of the otherwise cleverly designed theoretical algorithm. These points will be discussed in the following section.

\section{Implementation Insights \\ and Hurdles in the NISQ-era}
\label{sec_insights}


Besides problems that can occur when executing circuits on today's NISQ devices, we have used the presented QBM algorithm as an example to work out points that can lead to issues when transferring a quantum algorithm into an executable implementation.
In addition to that, in the rapidly growing domain of quantum software stacks, not all available SDKs support necessary features for an efficient implementation.
A combination of these points will be addressed in this section, combined with the perspective for what types of algorithms these points might be an obstacle and, if possible, how to mitigate some of these issues.

To ensure the verifiability of our results, we provide the source code of our implementation here~\cite{Paul2022_Impl}.
The code was written within the open-source quantum computing framework Qiskit~\cite{qiskit} at version 0.31.0.
The implementation was developed in the context of a quantum chemistry use case with the goal of approximating electronic ground states of molecules.

\subsection{Scaling of sampling quality}
\label{sec_shot_scaling}

In order to sample the wave function from the distribution, the circuit must be executed multiple times.
However, the sample size (i.e., the number of shots) should be the same order of magnitude as there are states to be represented.
Following this argument and assuming that for a given $n$-qubit Hamiltonian a large portion of the possible basis states contribute to the ground state, the number of required shots scales exponentially as $\mathcal{O}(2^n)$.
Thus, in the regime where a complex $2^n$-dimensional distribution might be difficult to access classically, an exponentially growing number of shots has to be performed.
Even for usual single-shot, small-depth circuit execution times of $\sim\mathcal{O}(\mu s)$ an exponentially growing number of repetitions might be a limiting factor.
As a reference, currently available NISQ-devices support a maximum of around 20,000 - 100,000 shots.
Even if technically possible to allow for a much larger number of shots, stability of the calibration of the underlying hardware must be ensured in order to get meaningful results.
If this is the case, an efficient generation of the probability distribution on the quantum register is possible at the expense of many circuit executions in order to sample it.
This, however, is not just an limiting aspect of the discussed QBM algorithm but is rather an issue for any quantum algorithm that relies on sampling for representing a distribution of states encoded in a quantum register.



\subsection{Classical Post Processing}
\label{sec_post_processing}

Conventional variational approaches like, e.g., the Variational Quantum Eigensolver (VQE)~\cite{vqe_original_paper1, vqe_original_paper2}, encode the target wave function and the problem Hamiltonian onto the quantum hardware in the form of quantum logic gates and allow, e.g., for the efficient evaluation of expectation values.
In contrast, the QBM approach generates a probability distribution as a basis for classically calculating the target wave function by summing over hidden layer configurations for a given visible layer configuration.
In order to evaluate expectation values necessary for the parameter optimization, the action of the problem Hamiltonian on the wave function must be calculated classically as well.
This essentially results in calculating the action of exponentially large (yet sparse) matrices on state vectors which ,especially in the context of computing expectation values, can be efficiently performed by quantum computers.
However, this could be a common problem for algorithms, which need to further process information about quantum states after it has been transferred to classical data.
Thus, it is important to be aware of additional classical calculations after the actual quantum computation in order to not lose the gained advantage by introducing a quantum step. 


\subsection{Mid-Circuit measurements}
\label{sec_mid_circuit_measurement}

As described in~\Cref{sec_quadratic_terms}, besides the data qubits from the visible and hidden registers, additional ancillary qubits are necessary in order to ensure a successful sampling. Naively, for $n$ qubits in the visible and $m$ qubits in the hidden layer, it requires additional $nm$ ancillary qubits to generate the probability distribution.
In this scenario, all measurements of the data and ancillary qubits can be performed at the end of the circuit, as it is usual for most circuits.
But by reusing a single ancillary qubit for all connections between visible and hidden layers, the required qubit resources reduce from $\mathcal{O}(n^2)$ to $\mathcal{O}(n)$.
This, however, requires both, the measurement and fast, reliable relaxation of the ancillary during the execution of the circuit.
Algorithmically, neither the relaxation nor the mid-circuit measurement pose a problem.
Including these features from a hardware and software perspective, however, is more challenging since they inherently affect the way quantum circuits and their execution results must be represented.
However, in order to efficiently implement the QBM, these features are necessary requirements for the hardware and software stack and are not yet supported by some, which limited the available options for the framework of choice.
As a step further, allowing for mid-circuit measurements would also open up the possibility of including conditional operations or operation layers on the register, based on the measurement results as, in principle, is intended in the discussed QBM algorithm as well.


\subsection{Ansatz universality}
\label{sec_ansatz_universality}

The ansatz for the target wave function given in~\eqref{eq_trial_wf} allows for the generation of real amplitudes with different signs for the basis states in order to approximate the ground state of the problem Hamiltonian.
Since the amplitudes of the ground state of an arbitrary Hamiltonian can be complex-valued, the ansatz itself does not allow for an arbitrary good approximation of the actual ground state.
As with many optimization problems, it is in fact difficult to compare the quality of the best known solution in the context of the method, with the globally optimal solution.
In order to address this issue, the originally proposed algorithm has been extended to allow for relative phases between basis states (see \cite{qbm_phase_node, qbm_ibmq_implementation_phase}). This can be done by including an imaginary part in the sign-node in~\eqref{eq_sign_node} according to
\begin{equation}
    s(v) = \tanh\left(\sum\nolimits_k (c_k + i\gamma_k)v_i + d + i\delta\right)~,
\label{eq_phase_node}
\end{equation}
with $\lbrace\gamma_k\rbrace, \delta$ being $n+1$ additional parameters.
But by using a phase-node, the covariant structure of the analytic gradient in~\eqref{eq_exp_val_derivative} is lost by then including real and imaginary parts of the involved quantities, thus not automatically cancelling out the shifts in~\eqref{eq_D_a}--\eqref{eq_D_w}.
Besides that, the sign- and phase-node pose another not directly apparent issue, which various Quantum Machine Learning models might struggle with.
This will be discussed next. 


\subsection{Ansatz expressivity}
\label{sec_ansatz_expressivity}

Now, even when assuming that the ground state of a given problem Hamiltonian has real-valued amplitudes, the algorithm proposed by Xia and Kais~\cite{Xia2018} would still not be able to approximate any ground state arbitrarily well.
This is due to the fact, that on the one hand, the most general real-valued $n$-qubit wave function has $2^n-1$ degrees of freedom -- one amplitude for each of the $2^n$ states minus one fixed amplitude due to normalization.
On the other hand, the number of parameters making up the model scales as $\mathcal{O}(n^2)$.
This \emph{information gap} becomes especially apparent for the expressivity of the sign-node:
Ideally, the node is supposed to realize relative signs between contributing states.
But since it is built from only $n+1$ parameters, it is not able to realize relative signs for all of the $2^n$ possible states.
For certain Hamiltonians this already posed a major issue for the solution quality obtained with as few as 2 qubits in the visible layer.
Note, that even by replacing the sign- with a phase-node, the limited expressivity is not resolved since it merely doubles the number of available parameters which can only contribute to the imaginary part of the amplitude.
In general, when building variational models, it is difficult to find a good compromise between expressivity and the number of parameters involved in building the model.
It is reasonable to assume that in the context of the QBM algorithm for a Hamiltonian, good approximations can only be found for the ground state if it is composed of only a few basis states and if there is only a small number of relative signs between basis states. 


\begin{figure}[tbp]
  \centering
  \subfloat[]{\includegraphics[scale=0.9]{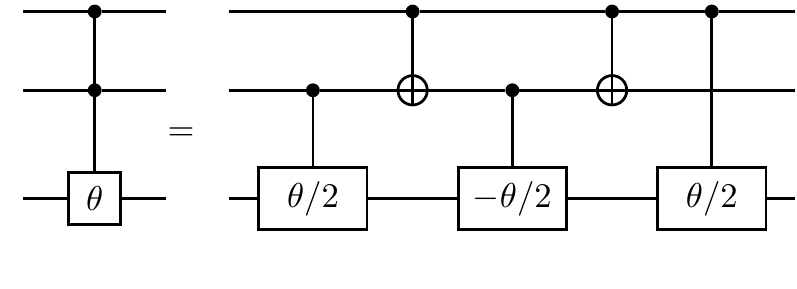} \label{fig_doubly_controlled_rotation_decomposition}} \\
  \subfloat[]{\includegraphics[scale=0.9]{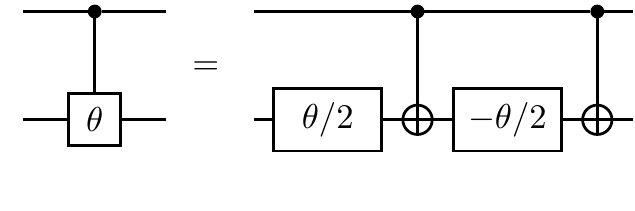} \label{fig_controlled_rotation_decomposition}}
  \caption{Decomposition of (a) a double-controlled rotation gate into controlled-rotation gates and CNOTs and (b) a controlled-rotation gate into one-qubit rotations and CNOTs. The gates can either be $R_z$- or $R_y$ rotations. All gates are $R_y$-rotation gates with the argument denoted in the box, although these decompositions hold for $R_x$- and $R_z$-rotations as well (for $R_x$-rotations replace the CNOTs in (b) with CZ-gates).} \label{fig_decompositions}
\end{figure}

\subsection{Width \& Depth}
\label{sec_width_depth}
As stated at the end of~\Cref{sec_optimization}, the width and depth requirements necessary for implementing the QBM algorithm scale as $\mathcal{O}(n^2)$ when assuming equally large visible and hidden layers (i.e., $m\sim\mathcal{O}(n)$).
At a first glance, this scaling seems fairly tolerable.
However, when implementing said algorithm and trying to execute it on current quantum devices, one might stumble over some points, which are not necessarily obvious:
For example, in the NISQ era, the prefactor of the depth scaling plays a non-negligible role.
Of course, in the context of complexity theory any prefactors and non-dominant terms can be neglected, but for implementing algorithms on today's NISQ-devices, e.g. just doubling the depth of an algorithm has a huge impact on the quality of the results.
This prefactor actually becomes quite large for the necessary double-controlled rotation gates when they are decomposed into gates, which can be executed on quantum backends.
This is necessary, because current backends support only a limited number of one- and two-qubit gates, into which any other gate described in an algorithm must be decomposed.

As a small example, assume that a backend supports $R_y$-gates and CNOTs. Following the decomposition rules of~\Cref{fig_doubly_controlled_rotation_decomposition} and~\Cref{fig_controlled_rotation_decomposition}, a single doubly-controlled rotation gate requires $8$ CNOT- and $6$ $R_y$-gates.
Thus, in order to implement all $n^2$ entangling layers necessary for generating the quadratic terms in the probability distribution, it actually requires $4n^2(2+3\cdot 2)= 32n^2$ two-qubit gates and $4n^2(3\cdot 2)=24n^2$ one-qubit gates.

In addition to the gate decomposition, even more two-qubit SWAP-gates are necessary if the interacting qubits cannot be directly entangled due to the quantum processor's topology.
Referring to the work of Leymann and Barzen~\cite{Leymann2020_BitterTruth}, this example was intended to point out that the choice of a particular backend has a major impact on the circuit depth and, thus, the quality of the results.
By analyzing the structure of an algorithm and the features of available backends, it is possible to improve on the solution quality.


\section{Conclusion \& Outlook}
\label{sec_conclusion}

To gain insight into the practicability of theoretically formalized quantum algorithms on current NISQ devices, we have implemented an algorithm for a quantum Boltzmann machine (QBM) proposed by Xia and Kais~\cite{Xia2018} and systematically summarized obstacles and limitations we have faced in the process.
Thereby, we have identified discrepancies between the domain of theoretical algorithm design and the practical application of quantum algorithms for actual use cases on current quantum hardware.
The systematically presented obstacles, limitations, and initially identified mitigation ideas can guide algorithm developers and practitioners to apply and implement a QBM, as well as similar algorithms.
One of the key findings is the following: Besides issues when transferring quantum data into classical data, e.g., when sampling from a distribution stored on the quantum register, to further process this data, we pointed out the desirable support of quantum hardware and quantum software for mid-circuit measurements. 
In our opinion, this feature holds potential for novel quantum algorithms, especially when considering the interchangeability of whole circuit operations conditioned on (multiple) measurement results.
Based on the decomposition of gates necessary for the algorithm at hand, we argue that the choice of an appropriate backend supporting a favorable set of gates is crucial for reducing the circuit depth and improving on the quality of results.
In this context, we would like to draw attention to promising automated backend and implementation selection approaches based on algorithm and hardware properties, as described in the work by Salm et al.~\cite{nisq_analyzer}.
Whilst experimenting with the implementation of different Hamiltonians (as described in~\Cref{sec_ansatz_expressivity}), the question has come up, in which way it might be possible to estimate the solution quality based on properties of the Hamiltonian and the ansatz of the variational circuit.

Furthermore, we are going to share our implementation of the QBM via the collaborative quantum software platform PlanQK~\cite{Leymann2020_QuantumInTheCloud, planqk_platform} to present and discuss our findings with further developers, quantum algorithm experts, and scientists in the quantum algorithm and quantum software development community. 
We are eager to abstract and generalize the essence of our findings along with proven mitigation ideas into design patterns for quantum algorithms and contribute them to the body of knowledge on quantum pattern languages started by Weigold et al.~\cite{Weigold2021}.

\section*{Acknowledgements}
\hl{This work was partially funded by the project PlanQK (01MK20005N) supported by the German Federal Ministry for Economic Affairs and Climate Action.}

\bibliographystyle{IEEEtran}
\bibliography{main}

\end{document}